%% file: lgcp_power_dependencies.tex
\documentclass[preprint]{elsarticle}
\usepackage{amsmath}
\usepackage{amsfonts}
\usepackage{amssymb}
\usepackage{amsthm}

\begin{document}
\include{introduction}
\include{theory}
\include{results_discussion}
\include{conclusion}

\bibliographystyle{plain}
\bibliography{lgcp_power_dependencies}

\end{document}

%% file: introduction.tex
\section{Introduction}
Experiments utilizing Lee-Goldburg (LG) excitation to determine molecular structure are continuing to develop rapidly. Accurate measurement of internuclear H-X distances is important in determining not only molecular structure but also dynamics \cite{Sny, Hong, Yee, Zech}. In particular, the growing field of NMR crystallography depends on accurate nuclear distance determination. Distance measurement relies on the magnetic dipole coupling between nuclei. LG techniques remove homonuclear dipolar coupling while leaving heteronuclear dipole coupling intact but scaled. Heteronuclear distances are measured by observing oscillations in build-up curves from cross-polarization experiments utilizing LG excitation on the abundant, high gamma nucleus.

A variety of experiments since the first seperated local field experiment by Waugh \cite{Wau} have been developed which employ LG cross-polarization (LGCP) along with magic angle spinning (MAS). Regardless of the implementation of LG excitation, frequencies in the dipolar coupled data are scaled. The theoretical scale factor is stated in some work to be $\cos{\theta_m} = \frac{1}{\sqrt{3}} = 0.577$ \cite{LG, Van}. Reports of scale factors range from 0.5 to 0.82 \cite{Van, Coe, Ram}. Some of these values contrast strongly with the theoretical value. Such variations in scale factors have not been adequately addressed in the literature.

In this work, experimental results for frequency switched LGCP with TPPM decoupling for the N1-H1 bond in guanosine are compared with simulated results on an isolated N-H system using SPINEVOLUTION \cite{Vesh}. Scale factors are calculated from simulation results. Further simulations are performed varying offset frequency and cross-polarization powers to exhibit scale factor variation.

%% file: theory.tex
\section{Theory}

The MAS Hamiltonian for dipolar coupled spins $I$ and $S$ in the rf interaction frame is \cite{Rov, Mul}

\begin{align}
H_{T}^{rf} &= D(t)\cos{\theta_I}\cos{\theta_S}I_{z}S_{z} + \sum_{n = -2}^{2} \omega_{Dn}[\sin{\theta_I}\sin{\theta_S}\frac{1}{2}(e^{i(\Sigma_{\textnormal{eff}}-n\omega_r)t}I_{+}S_{+} \notag \\
           & {}+ e^{i(\Delta_{\textnormal{eff}}-n\omega_r)t}I_{+}S_{-} + e^{i(\Delta_{\textnormal{eff}}-n\omega_r)t}I_{-}S_{+} + e^{i(\Sigma_{\textnormal{eff}}-n\omega_r)t}I_{-}S_{-}) \notag \\
           & {}- \sin{\theta_I}\cos{\theta_S}(I_{\pm}S_{z}e^{i(\pm\omega_{I,\textnormal{eff}} - n\omega_{r})t} - \cos{\theta_I}\sin{\theta_S}I_{z}S_{\pm}e^{i(\pm\omega_{S,\textnormal{eff}} - n\omega_{r})t})] \notag
\end{align}
There are many terms in this relation which we will describe. In order to simplify considerations, we will only be concerned with parts of the Hamiltonian that survive after a time integration over a rotor period. $D(t) \propto e^{-i\omega_{r}t}$ where $\omega_r$ is the MAS frequency. Thus it will integrate out. So, we have

\begin{align} \label{eq:dip_ham}
H_{\int}^{rf} &= \sum_{n = -2}^{2} \omega_{Dn}[\sin{\theta_I}\sin{\theta_S}\frac{1}{2}(e^{i(\Sigma_{\textnormal{eff}}-n\omega_r)t}I_{+}S_{+} \notag \\
           & {}+ e^{i(\Delta_{\textnormal{eff}}-n\omega_r)t}I_{+}S_{-} + e^{i(\Delta_{\textnormal{eff}}-n\omega_r)t}I_{-}S_{+} + e^{i(\Sigma_{\textnormal{eff}}-n\omega_r)t}I_{-}S_{-}) \\
           & {}- \sin{\theta_I}\cos{\theta_S}I_{\pm}S_{z}e^{i(\pm\omega_{I,\textnormal{eff}} - n\omega_{r})t} - \cos{\theta_I}\sin{\theta_S}I_{z}S_{\pm}e^{i(\pm\omega_{S,\textnormal{eff}} - n\omega_{r})t}] \notag
\end{align}

Other terms may or may not integrate out, depending on parameters described in what follows. Of particular importance is $\omega_{Dn}$, since $\omega_{Dn} \propto \delta$ with $\delta$ the dipolar coupling constant, and $\delta \propto \frac{1}{r^3}$ which provides the means to determine the internuclear distance, $r$. We are interested in finding the factors that scale $\omega_{Dn}$. The scaling factors are then  $\sin{\theta_I}\sin{\theta_S}$, $\sin{\theta_I}\cos{\theta_S}$, and $\cos{\theta_I}\sin{\theta_S}$ where the angles are measured with respect to the static field $B_{0}$ and

\begin{align}
\theta_I &= \tan^{-1}{\frac{\omega_{B_{rf}I}}{\Omega_{I}}} \\
\theta_S &= \tan^{-1}{\frac{\omega_{B_{rf}S}}{\Omega_{S}}}
\end{align}
where $\omega_{B_{rf}I}$ and $\omega_{B_{rf}S}$ are the nutation frequencies for the spins and
\begin{align}
\Omega_I &= -(\gamma_{I}B_{0} - \omega_{I_{rf}}) \\
\Omega_S &= -(\gamma_{S}B_{0} - \omega_{S_{rf}})
\end{align}
$\gamma_{I}$ and $\gamma_{S}$ are the gyromagnetic ratios and $\omega_{I_{rf}}$ and $\omega_{S_{rf}}$ are the rf field frequencies for respective spins. Let us look at the important factors in the time dependent exponents of Eq. \ref{eq:dip_ham}. We have

\begin{align}
\omega_{I,\textnormal{eff}} &= \sqrt{\Omega_{I}^{2} + \omega_{B_{rf}I}^{2}} \\
\omega_{S,\textnormal{eff}} &= \sqrt{\Omega_{S}^{2} + \omega_{B_{rf}S}^{2}} \\
\Sigma_{\textnormal{eff}} &= \omega_{I\textnormal{eff}} + \omega_{S,\textnormal{eff}} \\
\Delta_{\textnormal{eff}} &= \omega_{I\textnormal{eff}} - \omega_{S,\textnormal{eff}}
\end{align}
When any of these factors are an integral number of the rotor frequency, the corresponding term will time integrate to zero. Only when the time integrals are non-zero do the terms contribute. It is obvious that the scale factor can vary widely depending on frequency offsets and applied field strengths.

For LGCP experiments, we will take the $I$ spins to be the abundant, high gamma spins, and they will therefore be subjected to the LG resonance offset. So $\theta_I = \theta_{MA}$ where $\theta_{MA}$ is the magic angle. The $S$ spins will be spin locked by the $S$ channel rf field. With the $S$ field on resonance, $\theta_S = \frac{\pi}{2}$. To achieve a scale factor of 0.577, we need $\omega_{S,\textnormal{eff}} = n\omega_{r}$ for $n = -2,...,2$ and the other terms integrating to zero.

A peak resulting from zero quantum transitions will occur at frequency \cite{Mul}

\begin{align} \label{eq:mul}
q &= \frac{1}{2}((\omega_{I\textnormal{eff}} + \omega_{S,\textnormal{eff}})^2 + (\frac{1}{2}\delta \sin{\theta_I}\sin{\theta_S})^2)^{\frac{1}{2}}
\end{align}
when at least one of the rf fields is on resonance. For the LG conditions on $I$ and with the $S$ channel on resonance, $\delta$ is scaled by $\sin{\theta_I} = \sin{\theta_{MA}} = \sqrt{\frac{2}{3}}$. The first term is additive and corrupts the determination of $\delta$.


%% file: results_discussion.tex
\section{Results and Discussion}

\begin{figure}[ht]
 \centering
 \scalebox{.4}{\includegraphics{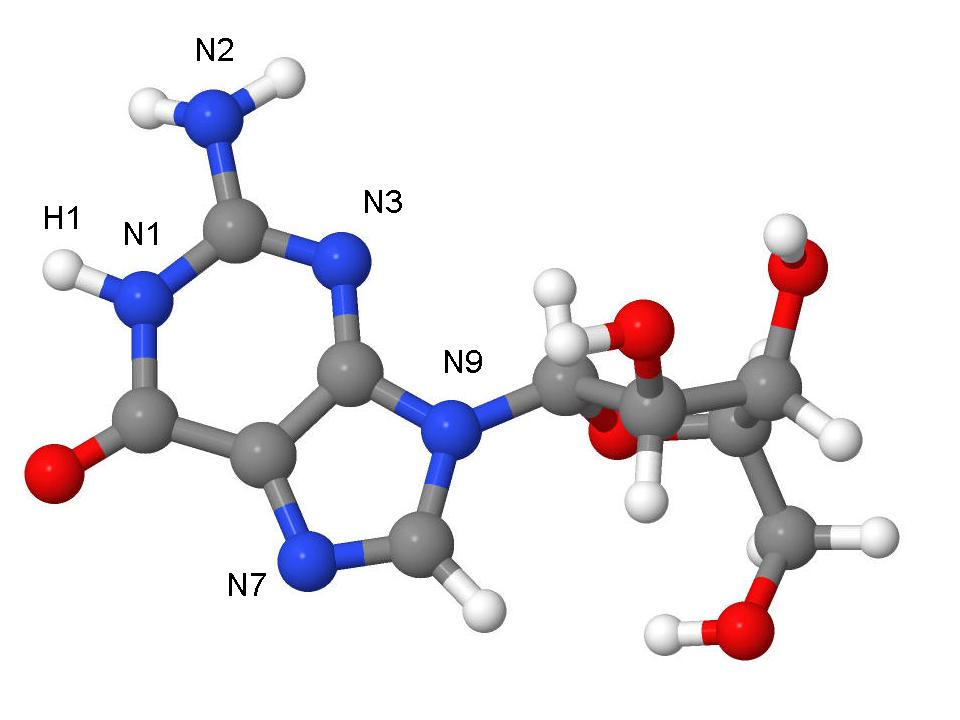}}\\
 \caption{N atoms and their associated H atoms in the guanosine molecule.}
 \label{fig:guan_Nlabel_nohilight}
\end{figure}

Experiments were performed on a Varian Infinity Plus 600 MHz magnet using a dual channel probe and 3.2 mm rotor. The guanosine sample enriched to 98\% $^{15}$N was purchased from Cambridge Isotopes. Figure \ref{fig:guan_Nlabel_nohilight} describes N positions in guansosine.
Prior to packing, the sample was placed in a humidity chamber for 14 days in order to ensure the dihydrate form of guanosine.

\begin{figure}[ht]
 \centering
 \scalebox{.4}{\includegraphics{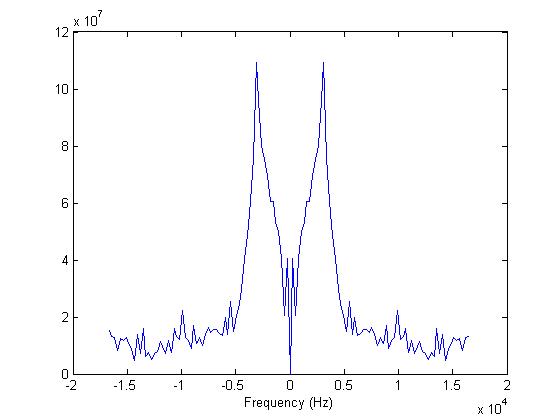}}\\
 \caption{Dipolar coupled data from a LGCP experiment. N and H rf powers are 53.05 and 50.00 kHz and LG offset frequency is 36.451 kHz.}
 \label{fig:guan_lgcpFH_p1_p865_p2_17p8_lg_19p4_ss10k}
\end{figure}

A frequency switched LGCP pulse sequence was used with TPPM decoupling and 10 kHz MAS. The sampling frequencies were 50 $\mu$s in the acquisition dimension and 30 $\mu$s in the evolution dimension. 128 points were taken in the evolution dimension which gives a frequency resolution of 260 Hz. The isotropic spectrum of guanosine has two lines for the N1 position. For this work, the upfield N1 line is chosen for all comparisons. The dipolar coupled data for N and H rf powers of 53.05 and 50.00 kHz and LG offset frequency of 36.451 kHz are shown in Figure \ref{fig:guan_lgcpFH_p1_p865_p2_17p8_lg_19p4_ss10k}.

\begin{figure}[ht]
 \centering
 \scalebox{.4}{\includegraphics{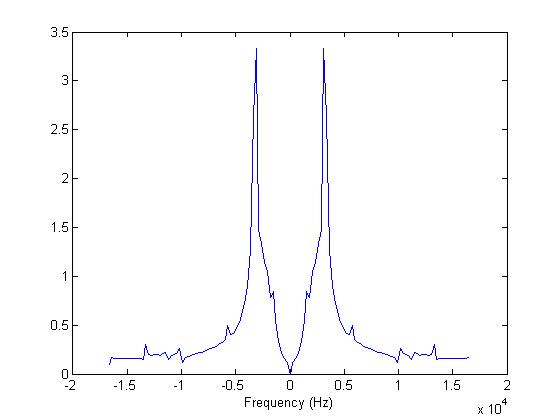}}\\
 \caption{Dipolar coupled data from a LGCP experiment. N and H rf powers are 53.05 and 50.00 kHz and LG offset frequency is 36.451 kHz.}
 \label{fig:p1_53p05_p2_50_foff_51p546_r_1p04_sim}
\end{figure}

A simple N-H system was used to simulate experiments. Other than H1, the closest hydrogen to N1 in guanosine is 2.43 $\mathring{A}$ away based on X-ray data \cite{Thew}. This gives a reasonably isolated N-H in the experiment which supports the simulation of an isolated N-H. Simulation parameters were set to match experimental powers, LG offset frequency, sampling frequency, and spinning speed.  N-H distance was varied to give the best match to experiment, giving a distance of 1.04 $\mathring{A}$. Figure \ref{fig:p1_53p05_p2_50_foff_51p546_r_1p04_sim} shows the simulation data for the LGCP experiment of Figure \ref{fig:guan_lgcpFH_p1_p865_p2_17p8_lg_19p4_ss10k}.
A distance of 1.04 $\mathring{A}$ gives a dipolar coupling of 10.823 kHz. From experiment, the dipolar coupling is 6250 $\pm$ 0.130 kHz which gives a scale factor of 0.578 $\pm$ 0.012.

\begin{figure}[ht]
 \centering
 \scalebox{.4}{\includegraphics{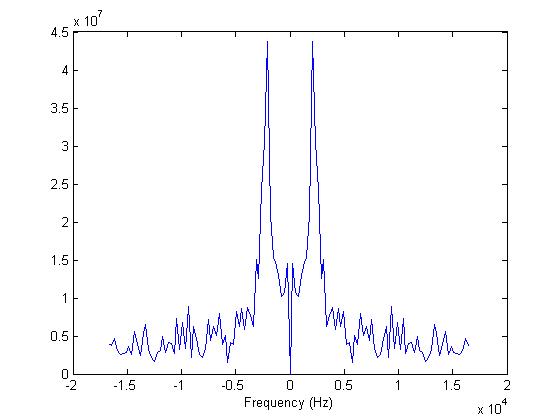}}\\
 \caption{Dipolar coupled data from a LGCP experiment. N and H rf powers are 58.3 and 63.13 kHz and LG offset frequency is 45.916 kHz.}
 \label{fig:guan_n15_lgcp_p1_p99_p2_15p4_lg_15p4_ss10k}
\end{figure}

\begin{figure}[ht]
 \centering
 \scalebox{.4}{\includegraphics{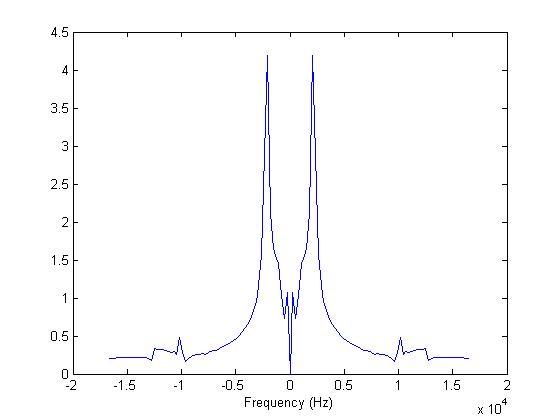}}\\
 \caption{Dipolar coupled data from a LGCP experiment. N and H rf powers are 58.3 and 63.13 kHz and LG offset frequency is 45.916 kHz.}
 \label{fig:p1_58p3_p2_63p13_foff_64p935_r_1p04_sim}
\end{figure}

Experimental and simulation data from a LGCP experiment with N and H rf powers of 58.3 and 63.13 kHz and a LG offset frequency of 45.916 kHz are shown in Figures \ref{fig:guan_n15_lgcp_p1_p99_p2_15p4_lg_15p4_ss10k} and \ref{fig:p1_58p3_p2_63p13_foff_64p935_r_1p04_sim}. In this case, the scale factor for the experiment is 0.385 $\pm$ 0.012. Simulation results for these experiments match qualitatively as well as quantitatively. It is shown in what follows that simulation results vary strongly with parameter settings.

\begin{figure}[ht]
 \centering
 \scalebox{.4}{\includegraphics{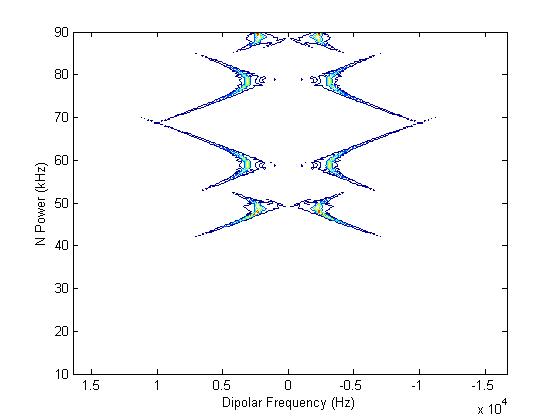}}\\
 \caption{Simulated dipolar coupled data as a function of N rf power from a LGCP experiment. H rf power is 56.18 and LG offset frequency is 39.725 kHz.}
 \label{fig:p2_56p18_foff_56p18_p1_10_p2_90_r_1p04}
\end{figure}

In order to investigate scale factor dependencies, SPINEVOLUTION was used to vary either N or H power while holding the LG offset frequency, N-H distance, and spinning speed constant. Figure \ref{fig:p2_56p18_foff_56p18_p1_10_p2_90_r_1p04} gives the contour plot of a simulated experiment varying the N rf power and with representative values of 56.18 kHz for H power and a 39.725 kHz LG offset. These values put the effective field for H at the magic angle in the rotating frame which is typically desired for LGCP to remove homonuclear dipolar coupling. 
It is seen from the plot that the peak follows a trajectory which will influence its scale factor. In fact, the simulation shows that for isolated N-H dipolar coupled systems, scale factors greater than 1 are possible. Of course, obtaining such results will depend on CP efficiency and probe capabilities.

Simulations varying H rf power were also run. As expected, H power affects the scale factor, since the angle at which the effective field is inclined to the z-axis in the rotating frame changes with H power. However, scale factors greater than 1 are also possible while holding the LG offset frequency fixed.

\begin{figure}[ht]
 \centering
 \scalebox{.4}{\includegraphics{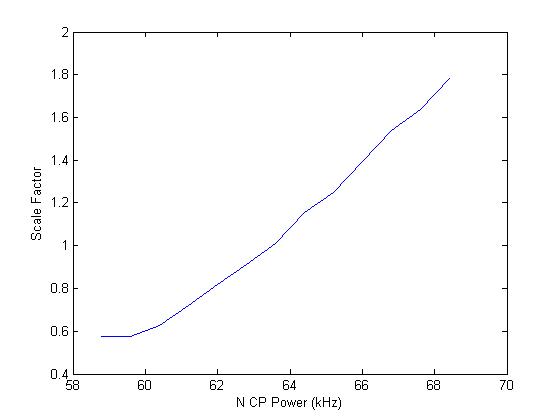}}\\
 \caption{Scale factor as a function of N CP power for a subset of data in Figure \ref{fig:p2_56p18_foff_56p18_p1_10_p2_90_r_1p04}}
 \label{fig:n_cp_power_vs_scale_factor}
\end{figure}

N CP power versus scale factor is plotted in Figure \ref{fig:n_cp_power_vs_scale_factor}. The plot shows a subset of the data shown in Figure \ref{fig:p2_56p18_foff_56p18_p1_10_p2_90_r_1p04}. The bumps in the plot at 64.6 and 66.8 kHz are from limits in the dipolar frequency resolution. As seen in the plot, the scale factor shows an non-linear increase with N CP power following Eq. \ref{eq:mul}.

%% file: conclusion.tex
\section{Conclusion}

It is shown in this work that SPINEVOLUTION simulations imply that scale factors in LGCP experiments can exhibit a range of values. The value of the scale factor is an interplay between rf field powers and frequency offsets. This work shows that in order to appropriately analyze LGCP results, simulations may be helpful. In addition, simulations might be used to tailor LGCP experimental parameters to give desirable scale factors which could aid in NMR crystallography.

%% file: lgcp_power_dependencies.bbl
\begin{thebibliography}{10}

\bibitem{Coe}
Coelho C, Rocha J, Madhu PK, and Mafra L.
\newblock Practical aspects of lee-goldburg based cramps techniques for
  high-resolution 1-h nmr spectroscopy in solids: implementation and
  applications.
\newblock {\em J Mag Res}, 194:264, 2008.

\bibitem{Hong}
Mei Hong, Xiaolan Yao, Karen Jakes, and Daniel Huster.
\newblock Investigation of molecular motions by lee-goldburg cross-polarization
  nmr spectroscopy.
\newblock {\em The Journal of Physical Chemistry B}, 106(29):7355--7364, 2002.

\bibitem{Wau}
Waugh JS.
\newblock Uncoupling of local field spectra in nuclear magnetic resonance:
  determination of atomic positions in solids.
\newblock {\em Proc Natl Acad Sci}, 73:1394, 1976.

\bibitem{LG}
Lee M and Goldburg W.
\newblock Nuclear magnetic resonance line narrowing by a rotating rf field.
\newblock {\em Phys Rev}, 140:A1261, 1965.

\bibitem{Vesh}
Veshtort M and Griffith RG.
\newblock Spinevolution: A powerful tool for simulating solid and liquid state
  nmr.
\newblock {\em J Mag Res}, 178:248, 2006.

\bibitem{Mul}
RR~Müller~L, Ernst.
\newblock Coherence transfer in the rotating frame.
\newblock {\em Molecular Physics}, 38(3):963--992, 1979.

\bibitem{Ram}
Opella~SJ Ramamoorthy~A, Wu~CH.
\newblock Experimental aspects of multidimensional solid-state nmr correlation
  spectroscopy.
\newblock {\em Journal of Magnetic Resonance}, 140(1):131 -- 140, 1999.

\bibitem{Rov}
D~Rovnyak.
\newblock Tutorial on analytic theory for cross-polarization in solid state
  nmr.
\newblock {\em Concepts in Magnetic Resonance Part A}, 32A(4):254–276, 2008.

\bibitem{Sny}
David~A. Snyder, Yang Chen, Natalia~G. Denissova, Thomas Acton, James~M.
  Aramini, Melissa Ciano, Richard Karlin, Jinfeng Liu, Philip Manor, P.~A.
  Rajan, Paolo Rossi, G.~V.~T. Swapna, Rong Xiao, Burkhard Rost, John Hunt, and
  Gaetano~T. Montelione.
\newblock Comparisons of nmr spectral quality and success in crystallization
  demonstrate that nmr and x-ray crystallography are complementary methods for
  small protein structure determination.
\newblock {\em Journal of the American Chemical Society}, 127(47):16505--16511,
  2005.

\bibitem{Thew}
Thewalt U, Bugg C, and Marsh R.
\newblock The crystal structure of guanosine dihydrate and inosine dihydrate.
\newblock {\em Acta Cryst}, B29:1089, 1970.

\bibitem{Van}
van Rossum~BJ, de~Groot~CP, Ladizhansky V, Vega S, and de~Groot~HJM.
\newblock A method for measuring heteronuclear distances in high speed mas nmr.
\newblock {\em J Am Chem Soc}, 122:3465, 2000.

\bibitem{Yee}
Adelinda~A. Yee, Alexei Savchenko, Alexandr Ignachenko, Jonathan Lukin, Xiaohui
  Xu, Tatiana Skarina, Elena Evdokimova, Cheng~Song Liu, Anthony Semesi,
  Valerie Guido, Aled~M. Edwards, and Cheryl~H. Arrowsmith.
\newblock Nmr and x-ray crystallography, complementary tools in structural
  proteomics of small proteins.
\newblock {\em Journal of the American Chemical Society}, 127(47):16512--16517,
  2005.

\bibitem{Zech}
Stephan~G. Zech, A.~Joshua Wand, and Ann~E. McDermott.
\newblock Protein structure determination by high-resolution solid-state nmr
  spectroscopy:? application to microcrystalline ubiquitin.
\newblock {\em Journal of the American Chemical Society}, 127(24):8618--8626,
  2005.

\end{thebibliography}
